\documentclass[reprint,superscriptaddress,amsmath,amssymb,prb]{revtex4-2}

\usepackage{graphicx} 
\usepackage{dcolumn}  
\usepackage{bm}       
\usepackage{xcolor}   
\usepackage{epstopdf}
\begin{document}

\preprint{APS/123-QED}
\title{Superconductivity in PtPb$_{4}$ with Possible Nontrivial Band Topology}

\author{C. Q. Xu}
\affiliation{School of Physics and Key Laboratory of MEMS of the Ministry of Education, Southeast University, Nanjing 211189, China}
\affiliation{Department of Physics and Astronomy, Michigan State University, East Lansing, Michigan 48824-2320, USA}

\author{B. Li}
\affiliation{Information Physics Research Center, Nanjing University of Posts and Telecommunications, Nanjing 210023, China}

\author{L. Zhang}
\affiliation{Department of Physics and Astronomy, Michigan State University, East Lansing, Michigan 48824-2320, USA}

\author{J. Pollanen}
\affiliation{Department of Physics and Astronomy, Michigan State University, East Lansing, Michigan 48824-2320, USA}

\author{X. L. Yi}
\affiliation{School of Physics and Key Laboratory of MEMS of the Ministry of Education, Southeast University, Nanjing 211189, China}

\author{X. Z. Xing}
\affiliation{School of Physics and Key Laboratory of MEMS of the Ministry of Education, Southeast University, Nanjing 211189, China}

\author{Y. Liu}
\affiliation{Key Laboratory of Quantum Precision Measurement of Zhejiang Province, Department of Applied Physics, Zhejiang University of Technology, Hangzhou 310023, China}

\author{J. H. Wang}
\affiliation{Wuhan National High Magnetic Field Center, School of Physics, Huazhong University of Science and Technology, Wuhan, 430074, China}

\author{Zengwei Zhu}
\affiliation{Wuhan National High Magnetic Field Center, School of Physics, Huazhong University of Science and Technology, Wuhan, 430074, China}

\author{Z. X. Shi}
\email{zxshi@seu.edu.cn}
\affiliation{School of Physics and Key Laboratory of MEMS of the Ministry of Education, Southeast University, Nanjing 211189, China}

\author{Xiaofeng Xu}
\email{xuxiaofeng@zjut.edu.cn}
\affiliation{Key Laboratory of Quantum Precision Measurement of Zhejiang Province, Department of Applied Physics, Zhejiang University of Technology, Hangzhou 310023, China}

\author{X. Ke}
\email{kexiangl@msu.edu}
\affiliation{Department of Physics and Astronomy, Michigan State University, East Lansing, Michigan 48824-2320, USA}

\date{\today}

\begin{abstract}

Superconductivity in topological quantum materials is much sought after as it represents the key avenue to searching for topological superconductors, which host a full pairing gap in the bulk but Majorana bound states at the surface. To date, however, the simultaneous realization of nontrivial band topology and superconductivity in the same material under ambient conditions remains rare. In this paper, we study both superconducting and topological properties of a binary compound PtPb$_{4}$ ($T_c$ $\sim$ 2.7 K) that was recently reported to exhibit large Rashba splitting, inherent to the heavy 5$d$ Pt and 6$p$ Pb. We show that in PtPb$_{4}$, the specific heat jump at $T_c$ reaches $\Delta C/\gamma T_{c}$$\sim$1.70$\pm0.04$, larger than 1.43 expected for the weak-coupling BCS superconductors. Moreover, the measurement of quantum oscillation suggests the possibility for a topological band structure, which is further studied by density functional theory calculations. Our study may stimulate future experimental and theoretical investigations in this intriguing material.

\end{abstract}

\maketitle

\section{Introduction}

Following the advent of topological insulators (TIs), the search for topological quantum materials with diverse symmetry-enforced topological states, characterized by the nontrivial topological invariants, has attracted a flurry of research interest in both condensed-matter physics and materials science\cite{Hasan,XiaoQi}. For instance, both three dimensional Dirac semimetals (DSMs) with fourfold degenerate band crossings and Weyl semimetals (WSMs) with pairs of Weyl nodes due to the broken inversion symmetry or time-reversal symmetry, have been demonstrated to exhibit extraordinary physical properties, such as extremely large magnetoresistance, ultrahigh carrier mobility, and chiral anomaly induced negative magnetoresistance\cite{Burkov,Armitage,Xiong,Qu,Ali-WTe2,Huang,Lv TaAs,Shekhar,Zhao Cd3As2,Narayanan Cd3As2,Li Cd3As2,Xiaolei Liu}. Another prominent example is the topological nodal-line semimetals (NLSMs) in which Dirac bands cross along a one-dimensional trajectory (line or loop) in momentum space, in contrast to the isolated Dirac nodes in TIs and DSMs. Thus far, topological nodal line states have been theoretically predicted and experimentally validated in some compounds, such as ZrSiS\cite{Ali ZrSiS,Schoop ZrSiS,Neupane ZrSiS,Fu ZrSiS}, PbTaSe$_{2}$\cite{Bian PbTaSe2,Guan PbTaSe2}, PtSn$_{4}$\cite{Wu PtSn4,Li PtSn4} etc.

Among these topological materials, those showing superconductivity at low temperatures are particularly intriguing since they provide arguably a natural platform to realize Majorana fermions, the manipulation of which forms the basis of future topological quantum computing. Majorana fermions are exotic particles which represent their own antiparticle and can arise in some novel physical systems, e.g., in the vortex core of topological superconductors\cite{L. Fu}. Arguably, superconductors with odd-parity pairing (e.g, UPt$_{3}$\cite{UPt3-1}, Cu$_x$Bi$_2$Se$_3$\cite{L. Fu}) or broken time-reversal symmetry (e.g., chiral LaPt$_3$P\cite{LaPt3P}) are strong candidates of topological superconductors\cite{Sato topological superconductors}. Nevertheless, these odd-parity or chiral superconductors are either rare or sensitive to disorder, and as such the ambiguous evidence for topological superconductivity is still elusive. Another feasible route to realize topological superconductors is to search for intrinsic superconductivity in topological materials, either in stoichiometric compounds at ambient or under high pressure, or by doping the topological materials to tune superconductivity. Indeed, superconductivity achieved in this manner has been observed in Cu$_x$Bi$_2$Se$_3$\cite{L. Fu,P. Das CuxBi2Se3,K. Matano CuxBi2Se3}, Au$_{2}$Pb\cite{Au2Pb-1,Au2Pb-2}, PbTaSe$_{2}$\cite{PbTaSe2 Le}, $\beta$-PdBi$_{2}$\cite{PdBi2-1,PdBi2-2}, S-doped MoTe$_2$\cite{S-MoTe2} etc., and in some cases, the experimental evidence of Majorana zero modes has been claimed.


In this broad context, the binary PtPb$_{4}$, a homologue of the Dirac nodal arc semimetal PtSn$_{4}$, was recently reported to exhibit large Rashba splitting around the $\Gamma$ point by ARPES\cite{PtPb4 Rashba}, which is inherent to the large atomic number of both Pt and Pb. PtPb$_{4}$ is particularly interesting as it also superconducts below $T_c$ $\sim$ 2.7 K \cite{PtPb4 Rashba,PtPb4 Gendron}, whereupon it is likely to be a candidate of topological superconductor. Despite the superconductivity being reported a few decades ago, a detailed study of its superconducting properties, in particular its association with putative topological states, is still lacking.

In this article, we report the synthesis of high quality PtPb$_{4}$ single crystal and the detailed study of its transport and thermodynamic properties by means of resistivity, magnetization and heat capacity measurements. We study the superconducting properties of PtPb$_{4}$ by measuring its lower and upper critical fields and investigate the possible topological properties by the de Haas-van Alphen (dHvA) oscillations and the first-principles calculations.

\section{Experimental details}

Large PtPb$_{4}$ single crystals up to several millimeters long were synthesized by the self-flux method \cite{PtPb4 Rashba}. Accurately weighed high-purity platinum pieces and lead shots were mixed with the molar ratio of 11 : 89. The mixed materials were placed in an alumina crucible and sealed in an evacuated quartz tube. The quartz tube was subsequently loaded into a box-type furnace and heated to 700$^{\circ}$C over a period of 10 hours and then kept at this temperature for one day to ensure complete melting. Afterwards, the furnace temperature was gradually cooled to 310$^{\circ}$C (at a rate of 1$^{\circ}$C/h) at which temperature the quartz tube was removed from the furnace and was immediately centrifuged to get rid of the excess flux. The resulting shiny PtPb$_{4}$ single crystals, with the lateral dimensions of several millimeters (as shown in Fig. 1(a)), were obtained. In order to eliminate the contamination from the possible residual flux, we used freshly cleaved samples for all measurements.

Single-crystal x-ray diffraction measurements were conducted at room temperature using the Rigaku diffractometer to check the crystalline quality. The actual chemical composition of the as-grown single crystals was determined using a field-emission scanning electron microscope (Hitachi Regulus 8230) equipped with an Oxford Ultim Extreme energy-dispersive X-ray spectroscopy (EDS). The electrical resistivity, heat capacity and magneto-transport measurements were carried out in the Quantum Design Physical Property Measurement System(QD-PPMS), and the magnetic susceptibility was measured in the Quantum Design Magnetic Property Measurement System(QD-MPMS).

We performed the electronic structure calculations using the generalized gradient approximation (GGA)\cite{GGA} as implemented in the full-potential linearized augmented plane wave (FP$-$LAPW) code WIEN2K \cite{Wien2k}. The muffin tin radii were chosen to be 2.5 a.u.\ for all atoms. The plane-wave cutoff was defined by $R \cdot K_{max}=7.0$, where $R$ is the smallest atomic sphere radius in the unit cell and $K_{max}$ is the magnitude of the largest $K$ vector. To calculate the topological properties and Fermi surfaces, we projected the Hamiltonian onto a basis made of $d$ Pt-centered, and $p$ Pb-centered orbitals, for a total of 136 Wannier functions, by means of the WANNIER90 and WannierTools packages \cite{wannier1,wannier2}. The Fermi surfaces were generated using a dense K-point mesh of $26 \times 15 \times 27$  in the Brillouin Zone. The related angular dependence of the quantum oscillation frequencies was calculated using the SKEAF code\cite{P. M. C. Rourke}. Since both Pt and Pb are heavy elements, relativistic effects and spin-orbit coupling (SOC) were taken into account in all calculations.

\section{Results}

\begin{figure}
\includegraphics[width=9cm]{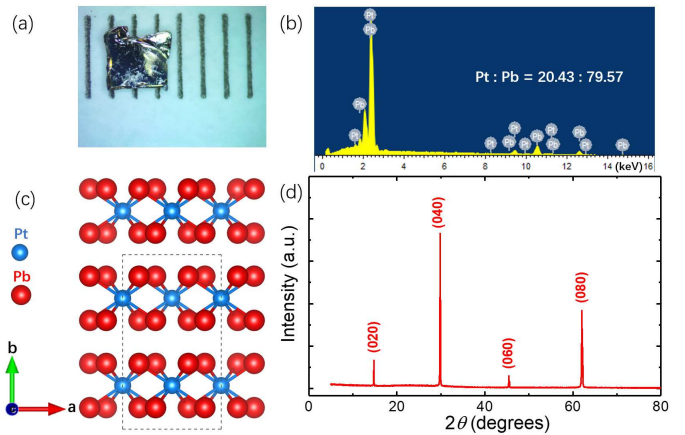}
\caption{\label{fig1}(a) The optical view of a single crystal used in this study. (b) The energy-dispersive X-ray spectroscopy (EDS) of a typical sample. (c) Schematic crystal structure of PtPb$_{4}$.  (d) X-ray diffraction pattern from the basal plane of a cleaved crystal, showing only (0 $l$ 0) reflections.}
\end{figure}

Figure 1(b) shows the typical EDS result measured on an as-grown PtPb$_{4}$ single crystal. The actual chemical stoichiometry is determined to be Pt : Pb = 20.43 : 79.57, very close to the nominal composition. Fig. 1(c) depicts the perspective of the crystal structure viewed along the $c$-axis, suggesting its layered structure. While PtPb$_{4}$ was previously reported to crystalize in the \textit{tetragonal} structure with the space group P4/nbm (No. 125), Lee \textit{et al.} recently found that it actually has the \textit{orthorhombic} lattice structure with the space group $Ccca$ (No. 68) by carrying out synchrotron x-ray diffraction study. It is conceivable that our crystals have the same structure with Lee's since almost the same procedures were used in the crystal growth. Following this newly determined orthorhombic crystal structure of PtPb$_{4}$ (i.e., the same as its sister compound PtSn$_{4}$), we present single crystal diffraction data in Fig. 1(d), where only (0 $l$ 0) peaks are observable and no extra impurity phases can be detectable, confirming high quality of our crystals.

\begin{figure}
\includegraphics[width=9cm]{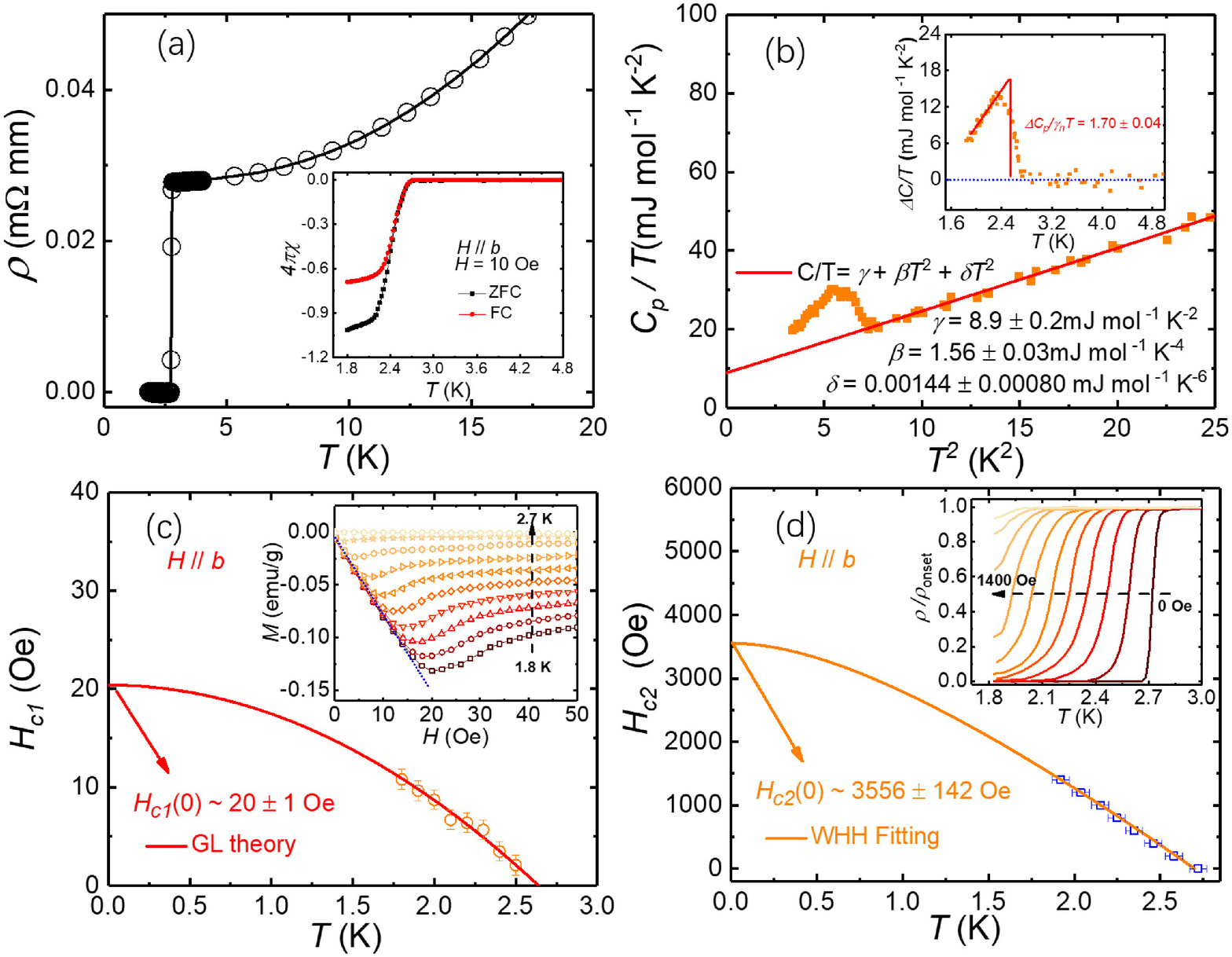}
\caption{\label{fig2} (a) $T$ dependence of resistivity measured under zero magnetic field. The inset shows the magnetic susceptibility as a function of temperature at a field of $H$ = 10 Oe with Zero-field-cooling(ZFC) and Field-cooling(FC) modes, respectively. (b) The heat capacity $C_{p}$ divided by temperature $T$ as a function of $T^{2}$ at low temperatures. The solid red line shows the fit to the data described in the main text. The inset shows $\Delta C/T$ versus $T$, where $\Delta C = C - C_{n}$, with $C_{n}$ being the heat capacity of the normal state. (c) The temperature dependence of the lower critical field $H_{c1}$ extracted from the inset. The red solid line is the fit using GL formula. (d) $T$-dependence of upper critical field $H_{c2}$ and its fit with the WHH formula. The inset shows the resistivity measured at various magnetic fields.}

\end{figure}

Fig. 2(a) illustrates the temperature ($T$) dependence of zero-field resistivity, with the inset showing the magnetic susceptibility under both zero-field-cooled (ZFC) and field-cooled (FC) protocols. It is notable that PtPb$_{4}$ exhibits metallic behaviors at high temperatures and undergoes a sharp superconducting transition below the critical temperature $T_{c} \sim 2.7$ K, in good agreement with the previous report\cite{PtPb4 Gendron}. The residual resistivity ratio(RRR), $\rho_{300K}/\rho_{3K}$, is approximately 50, indicative of high degree of metallicity. Both ZFC and FC magnetization curves show large diamagnetic susceptibility upon the superconducting transition. Bulk superconductivity is also evident from the low-$T$ heat capacity data measured at zero magnetic field (Fig. 2(b)), where a $\lambda$-shaped anomaly is clearly observed at $T_{c}$. The red solid curve in panel (b) represents the fit to the normal state heat capacity and its extrapolation to low temperatures. The normal state heat capacity is modeled with

\begin{eqnarray}
{C}_{p}/T = \gamma_{n} + \beta T^{2} + \delta T^{4},
\end{eqnarray}

\noindent where the first term is the electronic term and the other two are phononic contributions. The fit yields $\gamma_{n} = (8.9\pm0.2)$\,mJ/mol\,K$^{2}$, $\beta = (1.56\pm0.03)$\,mJ/mol\,K$^{4}$, and $\delta = (0.00144\pm0.0008)$\,mJ/mol\,K$^{6}$. The Debye temperature $\Theta_{D}$ is determined from the formula:

\begin{eqnarray}
\Theta_{D}=(\frac{12\pi^{4}}{5\beta}nR)^{1/3},
\end{eqnarray}

\noindent where $n$ is the number of atoms per formula unit, and $R$ = 8.31 J/mol K$^{-2}$ is the gas constant. The calculated $\Theta_{D}$ is (184$\pm5$) K. In the inset of Fig. 2(b), we present the heat capacity anomaly $\Delta C/T$ after subtracting the normal-state background. The value $\Delta C/\gamma T_{c}$ is thus estimated as $\sim$1.70$\pm0.04$, larger than 1.43 that is expected for the weak-coupling BCS superconductors. This $\Delta C/\gamma T_{c}$ value possibly results from enhanced electron-phonon coupling. Once the electron-phonon mechanism is assumed, the electron-phonon coupling constant $\lambda_{e-p}$ can be independently evaluated via the modified McMillian formula\cite{J. P. Carbotte,J. F. Mercure}

\begin{eqnarray}
\frac{\Delta C}{\gamma T_{c}} = 1.43[1+53(\frac{T_{c}}{\omega_{ln}})^{2}\ln(\frac{\omega_{ln}}{3T_{c}})],
\end{eqnarray}

\begin{eqnarray}
1.2T_{c} = \omega_{ln}\exp[\frac{-1.04(1+\lambda)}{\lambda-\mu^{*}(1+0.62\lambda)}],
\end{eqnarray}


\noindent where $\omega_{ln}$ is the average phonon-frequency,  $\mu^{\star}$ is the Coulomb pseudopotential and has a typical value of $\sim$0.13. Based on $\Delta C/\gamma T_{c}$ value obtained above, we find $\lambda_{e-p}$ = 0.83 $\pm$ 0.04. 
Here both $\lambda_{e-p}$ and $\Delta C/\gamma T_{c}$ are slightly larger than in many BCS superconductors and are very close to the intermediately coupled superconductors (e.g., indium with $T_c$ = 3.4 K, $\lambda_{e-p}$ = 0.81)\cite{J. P. Carbotte}. Alternatively, we can also derive $\lambda_{e-p}$ using the McMillan equation:

\begin{eqnarray}
\lambda_{e-p} = \frac{1.04+\mu^{*}\ln(\Theta_{D}/1.45 T_{c})}{(1-0.62\mu^{*})\ln(\Theta_{D}/1.45T_{c})-1.04}.
\end{eqnarray}

\noindent With the Debye temperature obtained above, $\lambda_{e-p}$ is estimated to be 0.62. 

The lower critical field $H_{c1}$ with magnetic field applied along the $b$-axis (i.e., perpendicular to the basal plane) was studied by examining the $M$-$H$ curves measured below $T_c$. As depicted in the inset of Fig. 2(c), the magnetization shows linear dependence on the field below $H_{c1}$ and it deviates from the linear slope above $H_{c1}$, characteristic of a type-II superconductor. The extracted lower critical $H_{c1}$ is plotted in the main panel of Fig. 2(c). By fitting the data using the Ginzburg-Landau (GL) equation

\begin{eqnarray}
H_{c1}(T) =  H_{c1}(0)[1 - (\frac{T}{T_{c}})^{2}],
\end{eqnarray}
\noindent we estimate $H_{c1}(0)$ as $\sim($20$\pm$1) Oe at $T$ = 0 K. Using the expression $\mu_0H_{c1}$=$\frac{\phi_0}{4\pi\lambda^2}$$(ln\frac{\lambda}{\xi})$ and the coherence length $\xi$ derived below, the magnetic penetration depth is calculated to be $\lambda$ $\sim$ (476.0 $\pm$ 50.0) nm.

We also study the upper critical field $H_{c2}$ by measuring the resistivity under various applied fields. As demonstrated in the inset of Fig. 2(d), the superconducting transition progressively shifts to lower temperatures and the width of transition becomes broadened with increasing field. We determine the upper critical field $H_{c2}$ by taking the points where the resistivity drops to 50\% of its normal state value right above $T_{c}$. The thus-determined $H_{c2}$ as a function of temperature is presented in the main panel of Fig. 2(d), with the out-of-plane ($H$$\parallel$ $b$) field direction. The solid curve in Fig. 2(d) represents the fit using the Werthamer-Helfand-Hohenberg (WHH) formula\cite{WHH}. $H_{c2}(0)$ is thus estimated to be $\sim$ (3556 $\pm$ 142) Oe, which is much smaller than the Pauli limit, suggesting the orbital effect at play here. Given $\mu_0H_{c2}$=$\frac{\phi_0}{2\pi\xi^2}$, this gives the coherence length $\xi$ $\sim$ (30.44 $\pm$ 0.61) nm and the GL parameter $\kappa$ = $\frac{\lambda}{\xi}$ $\simeq$ 15.64 $\pm$ 1.96.

\begin{figure}
\includegraphics[width=9cm]{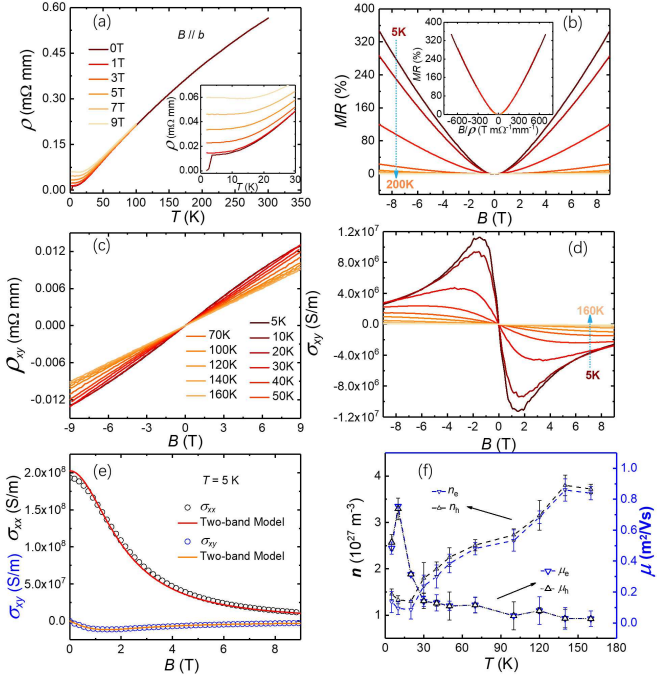}
\caption{Temperature dependence of resistivity at some selected fields. The inset shows an enlarged view of the low temperature region. (b) The MR at some fixed temperatures. The inset is the corresponding plot of Kohler's rule. (c) The Hall resistivity $\rho_{yx}$ at some selected temperatures. (d) The Hall conductivity ($\sigma_{xy}$) at the same temperatures as indicated in panel (c). (e) The fits of electrical conductivity $\sigma_{xx}$ and Hall conductivity $\sigma_{xy}$ by the two-band model at 5 K. (f) The carrier density and the mobility vs $T$ extracted based on the two-band fitting.}
\end{figure}

Fig. 3(a) shows the resistivity $\rho$ as a function of temperature in various magnetic fields. As seen, $\rho$(T) exhibits a weak upturn at low temperatures when the field is approaching 9 T. The magnetoresistance (MR), defined as MR=$\frac{\rho(H)-\rho(0)}{\rho(0)}\times 100\%$, has also been measured at various temperatures as shown in Fig. 3(b). The magnitude of MR ($\sim$350\% at 9 T and 5 K) does not show any sign of saturation. Interestingly, we find that, by plotting MR vs $B/\rho$, all curves collapse on each other as shown in the inset of Fig. 3(b), which suggests that the MR behavior of PtPb$_{4}$ follows the semiclassical Kohler's rule. Such a feature has also been observed in topological WTe$_{2}$\cite{WTe2 Kohler}, PdSn$_4$\cite{Lo PdSn4} etc. We also measured the Hall effect to determine the carrier types in PtPb$_{4}$. As a routine process, we antisymmetrize the data using $\rho_{xy}=\frac{\rho(+H)-\rho(-H)}{2}$. The Hall resistivity $\rho_{xy}$ is displayed in Fig. 3(c). As noted, the Hall resistivity shows nonlinearity with the field, consistent with the multiband features as evidenced from both the quantum oscillations and Fermi surface calculations to be discussed later. Then, we obtain the conductivity tensors $\sigma_{xx}$ and $\sigma_{xy}$ from the measured resistivity and Hall resistivity\cite{G. Grosso,J.Xu}:

\begin{eqnarray}
\sigma_{xx} = \frac{\rho_{xx}}{\rho_{xx}\rho_{xx}+\rho_{xy}\rho_{xy}}, \sigma_{xy} = - \frac{\rho_{xy}}{\rho_{xx}\rho_{xx}+\rho_{xy}\rho_{xy}}.
\end{eqnarray}

\noindent Afterward, we fit both the longitudinal ($\sigma_{xx}$) and the transverse magnetoconductivity ($\sigma_{xy}$) using the two-band model as parameterized below and the fits are illustrated in Fig. 3(e).

\begin{eqnarray}
\sigma_{xx} = -e(\frac{n_{e}\mu_{e}}{1+\mu_{e}^{2}B^{2}}+\frac{n_{h}\mu_{h}}{1+\mu_{h}^{2}B^{2}}),
\end{eqnarray}

\begin{eqnarray}
\sigma_{xy} = eB(\frac{n_{e}\mu_{e}^{2}}{1+\mu_{e}^{2}B^{2}}-\frac{n_{h}\mu_{h}^{2}}{1+\mu_{h}^{2}B^{2}}).
\end{eqnarray}

\noindent Here $e$ is the elementary charge -1.602$\times$10$^{-19}$C, $n_{e}$, $n_{h}$ are the carrier density, and $\mu_{e}$, $\mu_{h}$ are the corresponding mobility, of electrons and holes, respectively. It is clearly seen that both $\sigma_{xx}$ and  $\sigma_{xy}$ can be well fitted using this two-band model in the whole field range studied. The fitting gives $n_{h} = (1.45 \pm 0.11) \times 10^{27}$ m$^{-3}$, $n_{e} = (1.28 \pm 0.13) \times 10^{27}$ m$^{-3} $, $\mu_{h}$ = $(0.52 \pm 0.05)$ m$^{2}$/Vs, $\mu_{e}$ = $(0.48 \pm 0.05)$ m$^{2}$/Vs at 5 K. However, given the number of free parameters used in the model, the good fitting shown here does not exclude other sets of parameters which may fit the data equally well.

\begin{figure*}
\includegraphics[width=15cm]{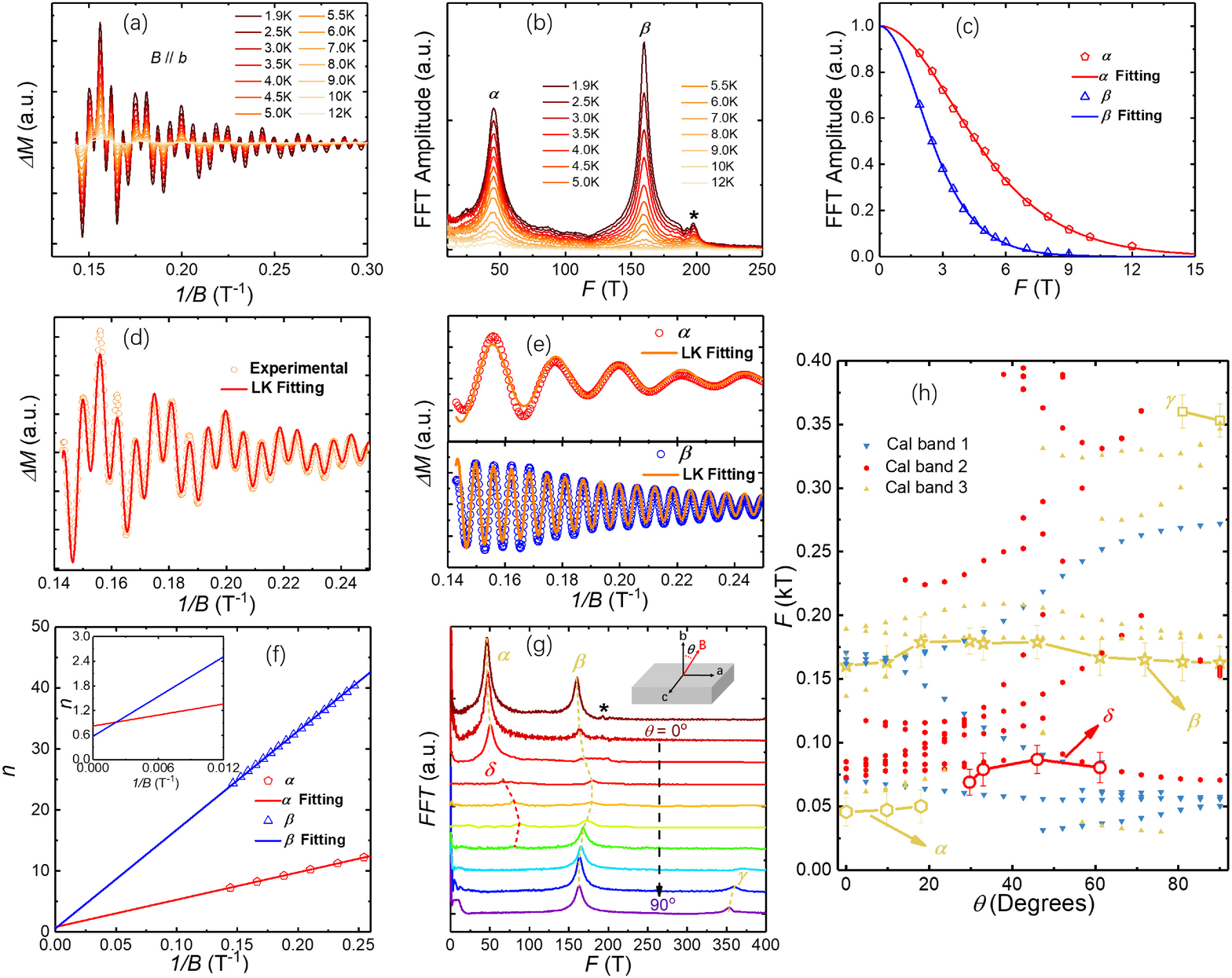}
\caption{(a) Oscillatory signals $\Delta$M vs magnetic field $B$ at some selected temperatures after subtracting the background. (b) FFT spectra of $\Delta$M oscillations. The asterisk marks the weak peak whose origin is not clear. (c) $T$ dependence of the FFT amplitude. The solid lines are fits to get the effective mass for each frequency. (d) The two-band LK fitting to the data measured at 1.9 K. (e) The $\alpha$ band and $\beta$ band oscillations were extracted using the FFT filter and fitted with the single-band LK formula. (f) The Landau's fan diagram for $\alpha$ and $\beta$. The inset enlarges the interception. (g) The angular dependence of FFT frequency at 1.9 K and the inset shows the rotation schematic diagram. (h) The angular dependence of FFT frequency from both experiment and calculations. The colors of $\alpha$, $\beta$, $\gamma$, $\delta$ bands are coded into their most-likely bands.}
\end{figure*}



%



To proceed, we study the electronic structure of PtPb$_{4}$ via quantum dHvA oscillations. The isothermal magnetization data with the magnetic field applied along the $b$ axis are measured. After subtracting the polynomial background, we show in Fig. 4(a) the oscillatory signals measured at different temperatures up to 12 K. The oscillations appear under a field as low as 2.5 T at 1.9 K, indicating good sample quality. The amplitude of oscillations decreases with increasing temperature. The fast Fourier transform (FFT) of the oscillatory magnetization $\Delta$$M$ shows two major frequencies, $F_{\alpha}$ = (45 $\pm$ 11) T, $F_{\beta}$ = (160 $\pm$ 9) T, 
shown in Fig. 4(b). Based on the Onsager relation $F=(\hbar/2\pi e)A_{e}$, we extracted the extremal Fermi-surface cross-sectional area $A_{e}$ associated with each frequency, for $\alpha$, $A_{e}$($\alpha$) = (0.43 $\pm$ 0.11)nm$^{-2}$, for $\beta$, $A_{e}$($\beta$) = (1.52 $\pm$ 0.09)nm$^{-2}$. Subsequently, we performed quantitative analysis of the oscillations using the standard Lifshitz-Kosevich(LK) formula\cite{Lifshits,Mikitik,Amit}:

\begin{equation}
\begin{aligned}
\Delta M \varpropto - R_{D}R_{T}\sin\{2\pi[\frac{F}{B}-(\frac{1}{2}-\phi)]\},
\end{aligned}
\end{equation}

\noindent where $R_{D}$ is the Dingle damping term ($R_{D}$=$\exp(2\pi^{2}k_{B}T_{D}m^{\ast}/eB\hbar$)), $R_{T}$ is the thermal damping factor ($R_{T}$=$\frac{2\pi^{2}k_{B}Tm^{\ast}/eB\hbar}{sinh(2\pi^{2}k_{B}Tm^{\ast}/eB\hbar)}$) and $\phi = \phi_{B}/2\pi - \delta$. Here $T_{D}$ is the Dingle temperature, $k_{B}$ is the Boltzmann constant, $m^{\ast}$ is the effective electron mass, $\phi_{B}$ is the Berry phase and $\delta$ is an additional phase shift that depends on the dimensionality of the Fermi surface, i.e., $\delta = 0$ for 2D Fermi surface, and $\delta = \pm\frac{1}{8}$ for 3D Fermi surface \cite{LuHZ}. As the Fermi surfaces of PtPb$_4$ are all 3D-like (see below), we only consider $\delta =\pm\frac{1}{8}$ here. As shown in Fig. 4(c), one can determine the effective mass $m^\ast$ for each frequency by fitting the temperature dependence of FFT amplitudes to $R_{T}$ and the fitting yields $m_{\alpha}$ = (0.15 $\pm$ 0.01) $m_{e}$, and $m_{\beta}$ = (0.27 $\pm$ 0.01) $m_{e}$ ($m_{e}$ is the bare electron mass). With the above fitted $m^\ast$, we can further fit $\Delta M$(1/B) to two-band LK formula at 1.9 K to obtain the corresponding Berry phase (see Fig. 4(d)). The fitting gives the Berry phase and Dingle temperature for each frequency as listed in Table I. For $\alpha$ band, if we assume $\delta = -\frac{1}{8}$, then $\phi_{B}= (1.35 \pm 0.08)\pi$, while $\delta = +\frac{1}{8}$, then $\phi_{B}= (1.85 \pm 0.11)\pi$. For $\beta$ band, the corresponding values are $\phi_{B}= (1.09 \pm 0.03)\pi$ ($\delta = -\frac{1}{8}$) and $\phi_{B}= (1.59 \pm 0.05)\pi$ ($\delta = +\frac{1}{8}$), respectively. On the other hand, by applying band filter, we also implement the single-band LK fitting for each mono-frequency oscillation, as shown in Fig. 4(e) for $\alpha$ and $\beta$-bands, respectively. Result from single-band fitting is nearly the same as that from two-band LK fitting (Table I). Alternatively, the Berry phase can also be estimated from Landau's fan diagram; as shown in Fig. 4(f), the valleys in the dHvA oscillation were assigned with Landau level index of n-1/4. From these linear fitting, we obtained $\phi_{B}=(1.39 \pm 0.17)\pi$ ($\phi_{B}=(1.89 \pm 0.23)\pi$) for $\delta = -\frac{1}{8}$ ($\delta = +\frac{1}{8}$) for $\alpha$ band, $\phi_{B}=(0.89 \pm 0.13)\pi$ ($\phi_{B}=(1.39 \pm 0.21)\pi$) for $\delta = -\frac{1}{8}$ ($\delta = +\frac{1}{8}$) for $\beta$-band, respectively, all of which are consistent with those from the LK Fitting. Although the preceding analysis seemingly points to the nontrivial Berry phase in this compound, it should be noted that this phase factor can be complicated by other contributions, such as the Maslov correction and the dynamic phase factor\cite{Alexandradinata PRX,Alexandradinata PRB,Clemens Schindler}.

\begin{table*}[t]
    \centering
    \caption{Parameters obtained from the dHvA oscillations}
    \begin{tabular}{cccccccccccccccc}
        \hline\hline

&&&&$H$ $\parallel b$ &&&$\phi_{B}$ \\
       & Methods & band & $F$(T) & $m^*/m_{e}$& $T_{D}$(K)& $A$(nm$^{-2}$)& $\delta=-\frac{1}{8}$ & $\delta=+\frac{1}{8}$  &    \\
       \hline
       &Two band LK fitting & $\alpha$ & (45 $\pm$ 11)& (0.15 $\pm$ 0.01)& (10.8 $\pm$ 0.65)& (0.43 $\pm$ 0.11)&(1.35 $\pm$ 0.08)$\pi$&(1.85 $\pm$ 0.11)$\pi$\\
       &  &  $\beta$ & (160 $\pm$ 9)& (0.27 $\pm$ 0.02)& (3.22 $\pm$ 0.21)& (1.52 $\pm$ 0.09)&(1.09 $\pm$ 0.03)$\pi$&(1.59 $\pm$ 0.05)$\pi$\\

       &One band LK fitting & $\alpha$& (45 $\pm$ 11)& (0.15 $\pm$ 0.01)& (9.04 $\pm$ 0.81)&(0.43 $\pm$ 0.11)&(1.37 $\pm$ 0.08)$\pi$&(1.87$\pm$ 0.11)$\pi$\\
       &  &  $\beta$& (160 $\pm$ 9)& (0.27 $\pm$ 0.02)& (2.74 $\pm$ 0.24) &(1.52 $\pm$ 0.09)&(1.08 $\pm$ 0.03)$\pi$&(1.58$\pm$ 0.05)$\pi$\\

       &Landau fan's diagram & $\alpha$& (45 $\pm$ 2)& (0.15 $\pm$ 0.01)&&(0.43 $\pm$ 0.11)&(1.39 $\pm$ 0.17)$\pi$&(1.89 $\pm$ 0.23)$\pi$\\
       &  &  $\beta$& (162 $\pm$ 3)& (0.27 $\pm$ 0.02)&&(1.53 $\pm$ 0.09)& (0.89 $\pm$ 0.13)$\pi$&(1.39 $\pm$ 0.21)$\pi$\\

        \hline\hline
    \end{tabular}

\end{table*}


Fermiology of PtPb$_{4}$ was further studied by the angular dependence of dHvA. From the angle-dependent FFT spectrum (Fig. 4(g)), the angle evolution of the frequency can be tracked. It is evident that both $\alpha$ and $\beta$ bands are quasi-3D with small anisotropy. We also found the third frequency $F_{\gamma}$= (353 $\pm$ 13) T when the field is applied along the $a$ axis, even though it is weak and fades away quickly with the field angle. Furthermore, to identify the orbital origin of each frequency, we investigated the angle dependence by the first-principles. As shown in Fig. 4(h), frequencies of $\alpha$ and $\beta$ obtained from experiment may be assigned to band 3 while $\delta$ band can be assigned to band 2, although outstanding discrepancies between experiment and the calculations remain.

\begin{figure*}
\includegraphics[width=16cm]{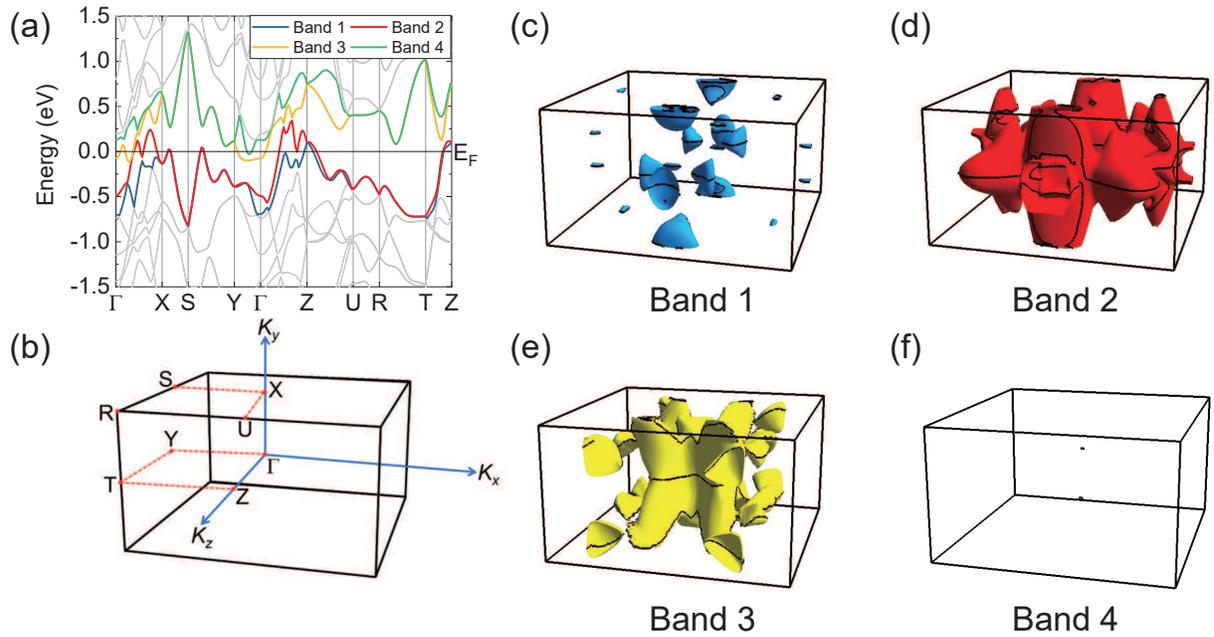}
\caption{\label{band}(a) Calculated band structure for orthorhombic PtPb$_4$ with SOC included. The bands crossing the Fermi level are marked by different colors and labeled by numbers. (b) Brillouin zone with the high-symmetry points specified. (c-f) 3D Fermi surfaces for each band. The extremal orbits are illustrated by lines for bands 1-3 when the magnetic field is applied along the $b$-axis.}
\end{figure*}

To further understand the electronic structure of PtPb$_{4}$, we perform the density functional theory calculations. Following the discussions in Ref. \cite{PtPb4 Rashba}, we used the \textit{orthorhombic} structure with centrosymmetric space group Ccca (No. 68) in the band structure calculations, with optimized lattice parameters $a$ = 6.733 {\AA}, $b$ = 11.625 {\AA}, and $c$ = 6.672 {\AA}. The conventional cell of the structure contains 16 Pb atoms and 4 Pt atoms, with stacked Pb-Pt-Pb layers along the $b$ axis. The calculated band structure with SOC is shown in Fig. \ref{band}(a). Four bands crossing the Fermi level are illustrated by different colors, where band 1 and 2 are hole-like, band 3 and 4 are electron-like. We note that the dispersions along  $X-S-Y$ and $U-R-T$ are highly degenerate for band 1 and band 2. It is the same for the dispersions of band 3 and band 4. We also note the hole-like bands and electron-like bands are nearly symmetrical with respect to the Fermi level, especially around the $S$ point. Fig. 5(b) shows the Brillouin zone with the high-symmetry points specified and Fig. 5(c)-(f) resolve the 3D Fermi surfaces for each band. The Fermi surfaces (FSs) of band 1 consist of several small pieces of Fermi pockets. For band 2, FSs consist of connected barrel-like Fermi pockets surrounding the Brillouin zone (BZ) center. The FSs of band 3 show a butterfly-like pocket in the BZ center with several small pockets around it, while FSs of band 4 are composed of two tiny dots. Our results are overall consistent with Lee's work\cite{PtPb4 Rashba}. Although the lattice parameters $a$ and $c$ are very close, the Fermi surfaces demonstrate clear anisotropy along $k_{x}$ and $k_{z}$ directions. To evaluate the possible topologically non-trivial states, we calculated the $\mathbb{Z}_2$ topological invariants on the six time-reversal invariant momentum planes using the Wannier charge centers (WCCs) calculations \cite{Soluyanov2011}, for both orthorhombic and tetragonal PtPb$_4$ (space group P4/nbm, No. 125) where the direct energy gap still appears at every $k$-point, allowing us to define the topological $\mathbb{Z}_2$ invariant. The $\mathbb{Z}_2$ invariants for orthorhombic PtPb$_4$ are (0; 101), which indicate that orthorhombic PtPb$_4$ exhibits nontrivial weak topology. As a comparison, the tetragonal PtPb$_4$ is found to be a strong topological insulator (TI) with $\mathbb{Z}_2$ (1; 000). A topological phase transition may occur in PtPb$_4$, accompanied by $C_4$ to $C_2$ symmetry breaking. In this sense, the intrinsic superconductivity in PtPb$_4$ ($T_c$ = 2.7 K) \cite{PtPb4 Gendron} may provide a natural platform to study the topological superconductivity or topological phase transition in the future.

\section{Conclusion}

In conclusion, we presented the detailed studies of the transport properties and quantum oscillations of a binary superconductor PtPb$_{4}$ that is isostructural to the NLSM PtSn$_{4}$. The heat capacity jump at $T_c$ is slightly enhanced compared with the weak coupling BCS superconductors. The normal state magnetotransport can overall be described by the two-band model. We also investigated its possible topological characteristics by the quantum oscillations and the DFT calculations. The possible nontrivial band topology proposed in this study, along with the large Rashba splitting observed in ARPES\cite{PtPb4 Rashba}, makes PtPb$_{4}$ a plausible venue to search for novel superconductivity and spintronic applications. Our findings in this paper provide a simple material platform to look into the potential of intriguing superconductivity and shall motivate more investigations on this material and its derivatives, both theoretically and experimentally.
\\

\begin{acknowledgments}

The authors would like to thank C. M. J. Andrew, Xianqing Lin, Pabitra Biswas, Sudeep Ghosh for useful discussions. This work is sponsored by the National Natural Science Foundation of China (Grant No. 11974061, No. 51861135104, No. 11574097, No. U1932217 and No. 12004337) and The National Key Research and Development Program of China (Grant No. 2018YFA0704300, No. 2016YFA0401704). X. K. acknowledges the financial support from the start-ups at Michigan State University. J. P. and L. Z. acknowledge the generous support of the Cowen Family Endowment at Michigan State University.

C. Q. Xu and B. Li contributed equally to this work.

\end{acknowledgments}

\appendix

\end{document}